# Field-effect chirality devices with Dirac semimetal

*Jiewei Chen, Ting Zhang, Jingli Wang, Ning Zhang, Wei Ji, Shuyun Zhou and Yang Chai\**

J. Chen, Dr. J. Wang, Dr. N. Zhang and Prof. Y. Chai*
Department of Applied Physics, The Hong Kong Polytechnic University, Hung Hom, Kowloon, Hong Kong, P. R. China
E-mail: ychai@polyu.edu.hk (Y.C.)

Prof. T. Zhang
Department of Physics, Hong Kong University of Science and Technology, Hong Kong, P. R. China

Prof. W. Ji
Department of Physics, Renmin University of China, Beijing 100872, P. R. China

Prof. S. Zhou
State Key Laboratory of Low Dimensional Quantum Physics, Department of Physics, Tsinghua University, Beijing, P. R. China

Prof. T. Zhang
Institute of Advanced Study, Hong Kong University of Science and Technology, Hong Kong, P. R. China

Dr. J. Wang, Dr. N. Zhang and Prof. Y. Chai*
The Hong Kong Polytechnic University Shenzhen Research Institute, 518057, Shenzhen, P. R. China

Charge-based field-effect transistors (FETs) greatly suffer from unavoidable carrier scattering and heat dissipation. In analogy to valley degree of freedom in semiconductors, chiral anomaly current in Weyl/Dirac semimetals is theoretically predicted to be nearly non-dissipative over long distances, but still lacks experimental ways to efficiently control its transport. Here we demonstrate field-effect chirality devices with Dirac semimetal $PtSe_2$, in which its Fermi level is close to the Dirac point in conduction band owing to intrinsic defects. The chiral anomaly is further corroborated with nonlocal valley transport measurement, which can also be effectively modulated by external fields, showing robust nonlocal valley transport with micrometer diffusion length. Similar to charge-based FETs, the chiral conductivity in $PtSe_2$ devices can be modulated by electrostatic gating with an ON/OFF ratio more than $10^3$. We also demonstrate basic logic functions in the devices with electric and magnetic fields as input signals.

## 1. Introduction

Conventional field-effect transistors (FETs) rely on charge transport in semiconductors, which inevitably results in heat dissipation induced by scattering charge carriers, and leads to ever-increasing power consumption in integrated circuits[1-2]. It is highly imperative to develop low-power devices by adopting the carrier transport mechanisms with non-dissipative characteristics, for example, non-charge-based information carrier (spin, valley, *etc*)[3-4]. Massless Weyl fermions have unusual transport properties that are resulted from chiral properties and the exotic trajectories against external perturbations[5-7]. Chirality of the Weyl fermion is defined by the sign of spin projection along the momentum direction (**Figure 1a**)[8]. Under the parallel electric field (***E***), chiral charge is pumped between two Weyl nodes with opposite chiralities under non-zero Berry curvature, leading to chiral current throughout the whole sample, so-called chiral anomaly. The pumped chiral charges can be compensated by inter-valley transport through quasi-momentum transfer (Figure S1, Supporting Information)[9-10]. In analogy to the valley degree of freedom in two-dimensional (2D) transition metal dichalcogenides, the chirality degree of freedom in topological semimetals can also potentially carry and encode information without heat dissipation[8, 11-13]. Because the chirality imbalance is closely related to the transfer of chirality between fermions and gauge fields with the global topology[12, 14], the topologically protected chiral current is dissipationless even under strong interactions over long distance (up to centimeter scale[15]), much longer than the transport of valley[16] or spin[17] (typical micrometer level). Compared with the successful modulation of valley/spin transport by optical or electrical stimulation[2, 16], the manipulation of chiral anomaly current in three-dimensional (3D) topological semimetals still remains elusive, because it lacks effective ways to modulate chiral anomaly current in semimetals.

In this work, we investigate the field-effect chiral anomaly characteristics by adopting Dirac semimetal $PtSe_2$, which possesses a 3D Dirac point in the conduction band. Electrostatic modulation with ionic liquid allows efficiently tuning the energy separation between the Dirac point and the Fermi level as well as the chiral conductivity. The chirality devices exhibit analogue characteristics of FETs with ON/OFF ratio higher than $10^3$ under external field modulation, and can perform basic logic functions. Chiral anomaly nonlocal valley transport can also be effectively modulated by both magnetic and electric fields. Our results provide another potential way for low-power electronics with new degree of freedom.

## 2. Results and discussion

### 2.1. Chirality-based field-effect device

Two Weyl nodes with distinct left-handed and right-handed chiralities are split from the Dirac cone under magnetic field (***B***), where the distance between two nodes in momentum space is determined by the magnitude of ***B***[7, 18]. The chemical potential imbalance between left-handed and right-handed chiralities drives the formation of chiral current with the presence of parallel ***E*** (Figure S2, Supporting Information), which results in chiral current, and exhibits negative longitudinal magnetoresistance (MR)[19-21]. We can manipulate chirality (valley-like degree of freedom) in semimetals by controlling chiral anomaly strength, which arises from non-zero Berry curvature ($\Omega$) and is closely related to the energy separation ($\Delta\varepsilon$) between Fermi level ($E_{Fermi}$) and Weyl/Dirac point (Figure S2a, Supporting Information). Therefore, chiral anomaly current can be efficiently tuned through field-effect shift of Fermi level of semimetals. When ***B*** // ***E***, the chiral anomaly conductivity ($\sigma_{chiral}$) can be described according to Equation (1)[22]:

$$\sigma_{chiral} = \frac{e^4 v_F^3 \tau B^2}{4\pi^2 \hbar \Delta\varepsilon^2} \qquad (1)$$

where $e$ is the electron charge, $v_F$ is the Fermi velocity near the Weyl points, $\tau$ is the inter-valley scattering time, and $\hbar$ is the Planck constant. It is noteworthy that this equation is only restricted to relatively low magnetic field according to theoretical analysis[23]. In experiments, most researchers validated the equation below 6T. The detailed analysis of chiral anomaly current in Dirac semimetals is summarized (Supporting Note I, and Figure S2b, Supporting Information).

Chiral current can be used for dissipationless information processing by field-effect modulation (**Figure 1a**). When $E_{Fermi}$ is close to the Weyl point, the larger difference ($|\mu^L - \mu^R| \gg 0$) in chemical potentials between left- ($\mu^L$) and right-handed ($\mu^R$) fermions results in strong chiral current ("ON" state); when $E_{Fermi}$ is away from the Weyl point, chiral anomaly strength decays remarkably, leading to $|\mu^L - \mu^R| \approx 0$ and negligible chiral current ("OFF" state). **Figure 1b** schematically illustrates the chirality-based field-effect device, where chiral current flows in the direction parallel to ***B*** and ***E***. The operations and characteristics of chirality-based devices are similar to those in charge-based FETs (**Figure 1c**). Electrostatic gating can tune the carrier density and $E_{Fermi}$ of Dirac semimetals, thus modulating *Δε* and chiral anomaly current in chirality-based devices (**Figure 1d**). To effectively modulate chiral anomaly current, it is quite necessary to choose topological materials with suitable *Δε*. If *Δε* is close to zero, Equation (1) becomes invalid [24]. If *Δε* is too large, $\sigma_{chiral}$ is too weak for the detection, and it is hard to efficiently manipulate the $\sigma_{chira}$ to achieve high ON/OFF value.

## 2.1. **Chiral anomaly in Dirac semimetal PtSe$_2$**

Layered PtSe$_2$ has been revealed with strongly layer-dependent bandgap, evolving from semiconducting to semimetallic characteristics with the increase of thickness (**Figure S3**, Supporting Information)[25-26]. According to density functional theory (DFT) calculations (**Figure S4**, Supporting Information), the conduction band of bulk PtSe$_2$ shows a cone around the point D (0, 0.32, 0.5) (in the units of 2π/a, 2π/a, 2π/c, respectively) along three different directions: in parallel with G-M path, perpendicular to G-M path, and along the $k_z$ direction (**Figure S4b-4d**, Supporting Information), different from the Dirac cone previously verified in the valence band of PtSe$_2$[18,19]. 3D band structures demonstrate that the bands are linearly dispersing across these nearly straight cones (**Figure 2a** and **Figure 2b**), resulting from *d* orbitals of Pt and *p* orbitals of Se. This fourfold degenerated point can be described by the 4×4 Dirac matrix, identified as the Dirac fermion. The Dirac point is positioned ~0.55 eV above the Fermi level in the conduction band of pristine bulk PtSe$_2$, which is quite different from the Dirac point in the valence band (with ~1.1 eV below the Fermi level) of PtSe$_2$ reported in previous works[27-28]. The Dirac point in valence band exhibits large *Δε*, which makes it extremely difficult to modulate chiral anomaly current. In this work, we focus on the investigation of the Dirac point in the conduction band of PtSe$_2$ to achieve field-effect chirality devices.

The presence of defects in materials can dramatically change the position of Fermi level. For example, the Fermi level of Dirac semimetal Na$_3$Bi can be lifted up by ~0.4 eV because of defects[20]. Energy dispersive X-ray spectrum of the PtSe$_2$ crystal (Figure S5, Supporting Information) reveals Se deficiency (roughly PtSe$_{1.93}$), which increases electron density and upshifts Fermi level close to the Dirac point in conduction band. We performed Hall measurements through a 6-terminal structure (sample #1 with ~12 nm thickness, Figure S6, Supporting Information), which shows *n*-type Hall resistance from 300 K to 2 K (**Figure 2c**). The carrier density (*n*) can be extracted according to $n = \frac{1}{R_H e}, R_H = \frac{Rd}{B}$, where $R_H$ is the Hall coefficient, *e* is the electron charge, *R* is tested Hall resistance and *d* is the sample thickness. The electron concentration of the PtSe$_2$ sample decrease from 3.15 ×10$^{21}$ cm$^{-3}$ ($n_{2D}$= 3.78×10$^{15}$ cm$^{-2}$) at 300 K to 8.50×10$^{20}$ cm$^{-3}$ ($n_{2D}$=1.02 ×10$^{15}$ cm$^{-2}$) at 2 K. Such high electron carrier density is repeatable in another Hall sample #2 (Figure S7, Supporting Information). Lv *et al.* reported that the measured Hall carrier density of 1.5×10$^{20}$ cm$^{-3}$ in Weyl semimetal WTe$_{1.98}$ can upshift the Fermi level by 60~120 meV[29]. The DFT calculation based on the measured carrier density shows that the upshifted Fermi level gives rise to small *Δε* (Supporting Note II and Figure S8, Supporting Information), which allows the observation of chiral anomaly conductivity in our samples without applying gate voltage.

Angle-dependent negative MR is a strong indicator of chiral anomaly in PtSe$_2$ devices. Other possible factors for negative MR, such as weak localization and current jetting, were carefully examined and excluded (Supporting Note III and Figure S9, Supporting Information). The crystalline orientation of PtSe$_2$ can be identified according to the angle between two straight edges[30]. We input current along the sharp zigzag edges of PtSe$_2$, and perform angle-dependent MR test by rotating the angle ($\theta$). Figure S6e shows the MR curve as a function of $\theta$, clearly exhibiting the angle-dependent MR behaviors. Sample #1 (~12 nm thickness) exhibits positive MR (21.16%) under **B**⊥**E** (Figure S6f, Supporting Information), and negative MR (-0.97%) when **B**∥**E** (**Figure 2d**). The chiral conductivity ($\sigma_{chiral}$) as a function of **B**$^2$ (the inset of **Figure 2d**) can be well fitted by a straight line, in good consistence with the chiral anomaly equation (1). Inset of **Figure 2e** shows a 4-terminal PtSe$_2$ (sample #3, ~13 nm thickness) for investigating chiral anomaly related angle-dependent MR behaviors. There is angle-dependent negative MR behaviors in sample #3 at 2K (**Figure 2e**), exhibiting the maximum negative MR value (-0.82%) when **B**∥**E**. The negative MR is gradually suppressed after rotating $\theta$ away from $\theta$ = 0° (**B**∥**E**), and becomes positive at $\theta$ > 2° under 9 T (**Figure 2f**). For relatively thick sample #4 (~15 nm), it also shows angle-dependent negative MR, where the MR decreases from 36% (**B**⊥**E**) to -0.26% (**B**∥**E**) (Figure S10, Supporting Information). It is noteworthy that the negative MR ratio of #4 under **B**∥**E** is smaller than #3 because of stronger positive MR background in thicker sample. The thickness-dependent MR behaviors under **B**∥**E** and **B**⊥**E** are summarized in Supporting Table 1. It is quite necessary to adopt the samples with suitable thickness to observe chiral anomaly (Supporting Note IV, Figure S11 and Figure S12, Supporting Information). We performed magnetotransport tests on more than 10 PtSe$_2$ samples (8-15 nm), which show reproducible negative MR characteristics induced by chiral anomaly (Figure S13, Supporting Information). It is noteworthy that thin PtSe$_2$ samples (*e.g.*, sample #1 in Figure 2d) with relatively weak background conductance provides a good platform for investigating field-effect chiral devices.

## 2.2. Field-effect chirality devices

As $\sigma_{chiral}$ is strongly affected by $\Delta\varepsilon$, electrostatic modulation can be used to change the Fermi level and chiral anomaly conductivity. Due to the high equivalent capacitance of ~10 mF cm$^{-2}$, ionic liquid has been reported for effectively modulating the carrier density in transition metal dichalcogenides, which can tune the carrier density by two orders of magnitude from $2 \times 10^{12}$ cm$^{-2}$ to $1.4 \times 10^{14}$ cm$^{-2}$ in MoS$_2$ [31]. By applying ionic liquid gating (Figure S14a, Supporting Information), ions are driven to the channel surface, forming ultrahigh electrical double layer capacitance. There is negligible damage on the samples and insignificant leakage current during the gating process (Figure S15, Supporting Information). Resistance-temperature measurements on sample #8 (Figure S16, Supporting Information) show that different gating voltage ($V_g$) can affect the electrical conductivity of the PtSe$_2$, suggesting the shift of carrier density and Fermi level.

For sample #9 (~8.5 nm), negative MR is remarkably suppressed under negative $V_g$, and is enhanced under positive $V_g$ (**Figure 3a**) due to the change of $\Delta\varepsilon$. A positive $V_g$ can upshift the Fermi level, while $V_g$ downwardly shifts the Fermi level (Figure S14b, Supporting Information). The curve of $\sigma_{chiral}$ as a function of $B$ under different $V_g$ (Figure S17, Supporting Information) shows parabolic shape. The linear relationship between $\sigma_{chiral}$ and $B^2$ (**Figure 3b**) fits well with the Equation (1). **Figure 3c** shows the chiral conductivity as a function of $V_g$, in analogy to the transfer curve of depletion-mode charge-based FETs (**Figure 1c**). The ON/OFF ratio of chiral conductivity reaches 780 with electrical gating (+2V vs. -2V) under 6T. Because of high electron carrier density of the semimetal PtSe$_2$ at $V_g = 0$ and relatively large $\Delta\varepsilon$, it is difficult to drive the $E_{Fermi}$ across the Weyl point, thus resulting in the saturated characteristics similar to the transfer curve of conventional FETs.

Due to the thinner thickness (~8 nm), the MR of sample #10 changes from nearly 0 ($V_g$ = -2V) to -6.97% ($V_g$ = +2V) (Figure S18, Supporting Information), exhibiting much stronger electrostatic modulation than sample #4. The $\sigma_{chiral}$ of sample #10 under different gating is also linearly fitted (Figure S19a, Supporting Information) and it can be modulated by both electric and magnetic field (Figure S19b, Supporting Information), with the ON/OFF ratio of $1.0 \times 10^3$ (electrical field, +2V vs. -2V, at 9T) and $1.1 \times 10^3$ (magnetic field, 2T vs. 9T, at $V_g$= 2V), respectively. For thick sample #11 (~90 nm), it is hard to tune the MR from positive to negative because of strong positive MR background (Figure S20, Supporting Information). At another angle $\theta$ = 10°, we can successfully achieve field-effect MR in sample #4 (Figure S21, Supporting Information) by combining $\theta$, $V_g$ and $\bm{B}$ to manipulate chiral anomaly.

Chirality-based devices can perform logic operations, like the charge-based FETs[32]. We can define $V_g$ and $\bm{B}$ as two inputs, and $\sigma_{chiral}$ as output. For the input$_1$ ($\bm{B}$), logic "0" and "1" are defined as $\bm{B} \leq 4.5$ T and $\bm{B} > 4.5$T, respectively; for the input$_2$ ($V_g$), logic "0" and "1" are defined as $V_g \leq 0$ V and $V_g > 0$ V, respectively. AND logic function (**Figure 3d**) can be realized on sample #9 (~8.5 nm), where the output "0" corresponds to $\sigma_{chiral} \leq 780$ S cm$^{-1}$, and "1" corresponds to $\sigma_{chiral} > 780$ S cm$^{-1}$. Only when both input$_1$ and input$_2$ are logic "1", the output is "1". For thicker sample #15 (~15 nm), OR logic function can be achieved (Figure S22a, Supporting Information). The $\theta$ can be used as a switch to convert the logic function from "OR" to "AND" (Figure S22b, Supporting Information). In addition to two-input AND and OR logic operations, AND-OR logic function with three inputs can also be achieved by combining $\theta$, $V_g$ and $\bm{B}$ as the input signals (Supporting Note V and Figure S22c, Supporting Information).

### 2.3. Field-effect nonlocal valley transport

Nonlocal valley transport can distinguish the negative MR from conventional MR anisotropy, and provide another strong indicator to corroborate chiral anomaly in topological semimetals. We performed nonlocal valley transport through the "H" configuration (**Figure 4a**), in which constant current is injected and a valley imbalance is generated under parallel ***E*** and ***B***. This valley polarized states will diffuse over the sample like spin, and can be detected in the detector region with sufficient long distance (μm level)[8, 10],[33]. Nonlocal valley transport strength can be quantitatively characterized by width-dependent ratio and length-dependent exponential decay. We fabricated the devices in the "H" configuration with the same length (2 μm) and different widths (**Figure 4b** and Figure S23a, Supporting Information) to characterize the width-dependent valley transport. Constant current is applied through terminal 1-2, while terminal 3-4 (2 μm width) and 5-6 (1 μm width) are used to detect the voltage. Magnetic field can effectively modulate the nonlocal resistance of the device (**Figure 4c**). The pure nonlocal resistance is extracted by removing background from stray current[34] (Supporting Note VI, Supporting Information) according to $R_{\text{valley}} = -k_{34}Re^{-L/L_v}$ for terminal 3-4, where $R_{\text{valley}}$ reveals the strength of valley signal, $R$ is the local resistance, $L_v$ is the valley diffusion length, and $k$ is a dimensionless coefficient. The width ratio of terminal 5-6 to 3-4 is 0.5. Therefore, the corresponding theoretical nonlocal valley ratio should be 0.50 (dash line). The measured ***B*** dependent $R_{56-\text{NL}}/R_{34-\text{NL}}$ ratio is close to this theoretical value (the inset of **Figure 4c**). We also fabricated the device with different lengths (**Figure 4d** and Figure S23b, Supporting Information). By applying constant current through terminal 1-2, we tested the voltage of terminal 3-4 (2 μm length), 5-6 (4 μm length) and 7-8 (7 μm length). With the increase of lateral length $L$ (the length between current injected terminal pair and nonlocal detecting pairs of terminals), the nonlocal resistance (**Figure 4e**) decays rapidly.

Valley transport strength coefficient $\alpha_{NL}$ is used to quantitatively describe the strength of the nonlocal response[8, 10]:

$$\alpha_{\text{NL}} = \left|\frac{R_{\text{NL}}}{R_{\text{L}}}\right| \propto e^{-\frac{L}{L_v}} \qquad (2)$$

where $L_v$ is the inter-valley scattering length. The semi-log plot of $\alpha_{NL}$ against the lateral length $L$ (**Figure 4f**) clearly shows that the tested $\alpha_{NL}$ can be fitted by line, in which $\alpha_{NL}$ decays exponentially as the lateral length $L$ increases from 2 μm, 4 μm to 7 μm, consistent with the valley transport equation (2). The valley signal is relatively strong even for the 4 μm length[8, 10].

In addition to magnetic field, electrical gating can also be applied to manipulate the nonlocal valley transport, which shows that the tested $\alpha_{NL}$ can be well tuned ant fitted by straight line under $V_g = \pm 2V$ (**Figure 4f**). The good linear fitting in the semi-log curve supports that the tested valley signals decay exponentially under $V_g = \pm 2V$, fitting well with the equation (2). Both magnetic and electric field can successfully tune the strength of valley transport in different device geometries.

## 3. Conclusion

In summary, we design and demonstrate a device structure for controlling chirality transport in Dirac semimetal PtSe$_2$. The chirality-based field-effect devices show efficient electrostatic control over chiral anomaly current with the ON/OFF ratio more than $10^3$, and realize basic logic functions. The nonlocal valley transport is width- and length-dependent, which can also be effectively modulated by magnetic and electrical fields. This study provides a way to control the transport of chirality degree of freedom, which has the potential for significantly reducing power consumption compared with state-of-the-art charge-based FETs.

## 4. Methods

Device fabrication: PtSe$_2$ crystal was purchased from HQ Graphene. Flakes were mechanically exfoliated onto the Si substrate with 300-nm-thick SiO$_2$. Electron beam lithography technique was used to define the pattern of metal electrodes. Metal contacts were prepared by sequential thermal evaporation of Cr (10 nm) and Au (80 nm) at the rate of 0.2 Å /s and 0.5 Å /s, respectively. As there is layer-depdent bandstructure in transition-metal dichalcogenides[26, 35], we have chosen thick enough samples to show bulk-like bandstructures. To observe the obvious angle dependent negative MR, the thicknesses of used PtSe$_2$ flakes were selected in the range of 8-15 nm, determined by Bruker Multimode atomic force microscopy. To prepare devices for nonlocal valley transport measurement, a relatively low beam current of 30 pA, 20 kV was adopted in focused ion beam process to avoid obvious damage to the sample.

Magnetotransport measurement: The temperature-dependent magnetotransport measurements were carried out by Physical Property Measurement System (PPMS) from Quantum Design. The direction of the magnetic field was reversed to correct the additional Hall (or resistive) voltage signals due to the misalignment of the voltage leads during the MR (or Hall resistivity) measurements. The ionic liquid DEME-TFSI can induce ~$10^{14}$ cm$^{-2}$ carriers to form the equivalent capacitance of ~10 mF/cm$^2$. A droplet of ionic liquid was used to cover the surface of PtSe$_2$ and the side gate electrode. Then, the sample was kept under high vacuum with 9×10$^{-8}$ Torr for 24 h. The side ionic liquid gate pattern was prepared much larger than PtSe$_2$ film for efficient gating. The initial gating temperature was set at 220 K, which is close to freezing point of DEME-TFSI. Gating at this temperature allows the effective gating and avoids damaging the sample.

Materials characterizations: Scanning electron microscope spectrum and energy dispersive X-ray spectroscopy were acquired by JEOL Model JSM-6490. Raman spectrum was performed with a Witec alpha300 R (laser source, 532nm).

First-principles calculations: We employ first principle simulation method to investigate the precise band structure of bulk PtSe$_2$ material. Quantum-Espresso code[36-38] was used to perform the calculation. We used projector-augmented-wave (PAW) method[39] to describe the atomic potential, and generalized-gradient-approximation (GGA) Perdew-Bruke-Ernzerhof[40] exchange-correlation functional. The energy cutoff for plane-waves was set to 80 Ryd to precisely describe the highly-localized d-orbitals in the system. The PtSe$_2$ crystal was firstly fully optimized under k-sampling density 16 × 16 × 12, and then band structure was calculated. To search for possible Dirac point in the whole Brillouin zone (BZ), we used Wannier90 code[41] to project the 16 × 16 × 12 band structure onto maximally-localized Wannier orbitals[42] and generate a tight-binding model. With this tight-binding model, it is possible to search for tens of thousands of k-points in the BZ. After the Dirac point in conduction band was identified by this tight-binding model, it was further verified by Quantum-Espresso.

**Acknowledgements**

This work was supported by Research Grant Council of Hong Kong (Grant No. N_PolyU540/17), the Hong Kong Polytechnic University (Grant No. 1-ZVGH), and Science, Technology and Innovation Commission of Shenzhen (JCYJ20180507183424383). Ting Zhang acknowledge Research Grant Council of Hong Kong (Grant Nos. 6300818 and R6015-18) and the Hong Kong University of Science and Technology (Grant No. R9501).

**Figures and captions**

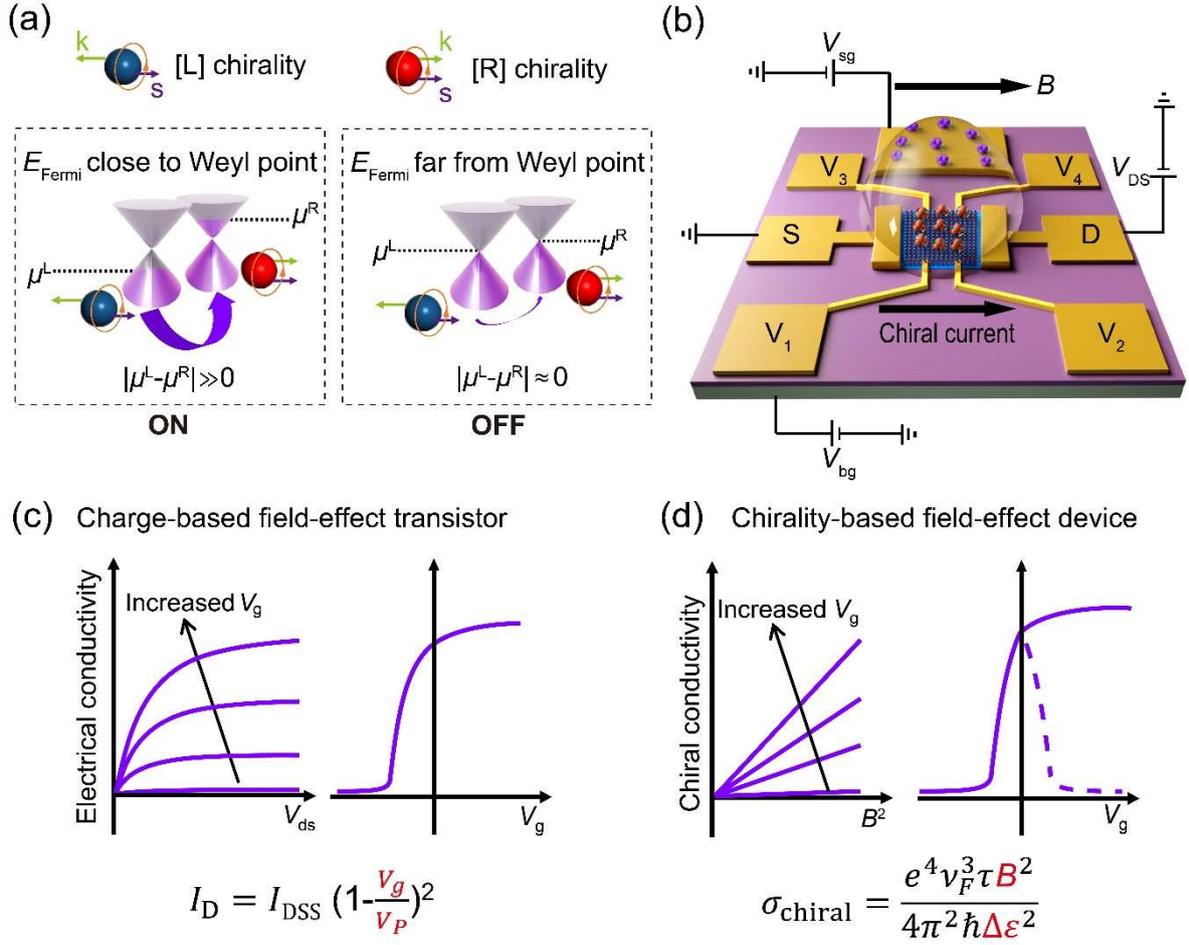

**Figure 1. Chirality-based field-effect devices.** (a) Chiral anomaly current in Dirac semimetal. "ON" state: $E_{\text{Fermi}}$ is close to the Weyl point and $|\mu^L - \mu^R| \gg 0$. "OFF" state: $E_{\text{Fermi}}$ away from the Weyl point and $|\mu^L - \mu^R| \approx 0$. (b) Illustration of device structure for chirality-based field-effect devices. (c) Schematic curves of electrical conductivity in charged-based field-effect transistor as a function of $V_{ds}$ (left panel) and $V_g$ (right panel), respectively. In an n-type depletion mode FET, $I_D$ is drain current, $I_{DSS}$ is maximum saturation current, and $V_P$ is the pinch-off voltage at which the channel closes. (d) Schematic curves of chiral conductivity in chirality-based field-effect devices as a function of $\boldsymbol{B}^2$ (left panel) and $V_g$ (right panel), respectively. For the curve as a function of $V_g$, the dash line refers to the situation that Fermi level can easily cross the topological point due to too small $\Delta\varepsilon$, while the solid line shows the case that Fermi level cannot cross the topological point.

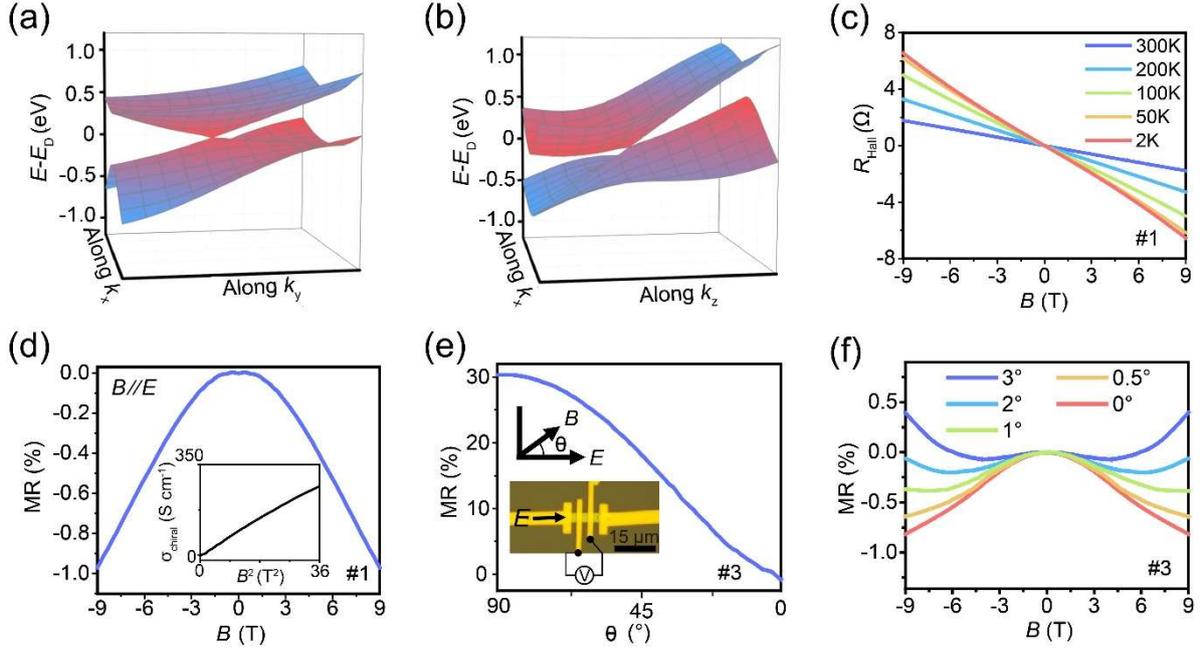

**Figure 2. Chiral anomaly in PtSe$_2$ semimetal.** (a) Three dimensional band structures along $k_x$ and $k_y$. (b) Three dimensional band structures along $k_x$ and $k_y$, where $k_x$ is the direction vertical to G-M line, and $k_y$ is the direction parallel to G-M line. The variable range is +/-0.1(unit 2π/a). (c) The Hall resistance varies from 300K to 2K of sample #1. (d) Magnetotransport behaviors of Hall bar device #1 under ***B//E*** at 2K. MR= $\frac{R(B)-R(0)}{R(0)} \times 100\%$, where $R(0)$ is the resistance at zero magnetic field, and $R(B)$ is the resistance under ***B***. The inset is σ$_{chiral}$ as a function of ***B***$^2$, extracted according to σ$_{chiral}$ = σ - σ$_0$, where σ is the conductivity under specific magnetic field, and σ$_0$ is the conductivity without magnetic field. (e) Angle-dependent longitudinal magnetic resistance of 4-terminal sample #3. Inset is the optical microscope photograph of typical tested 4-terminal sample. (f) Longitudinal MR of sample #3 at small rotation angles near ***B//E***.

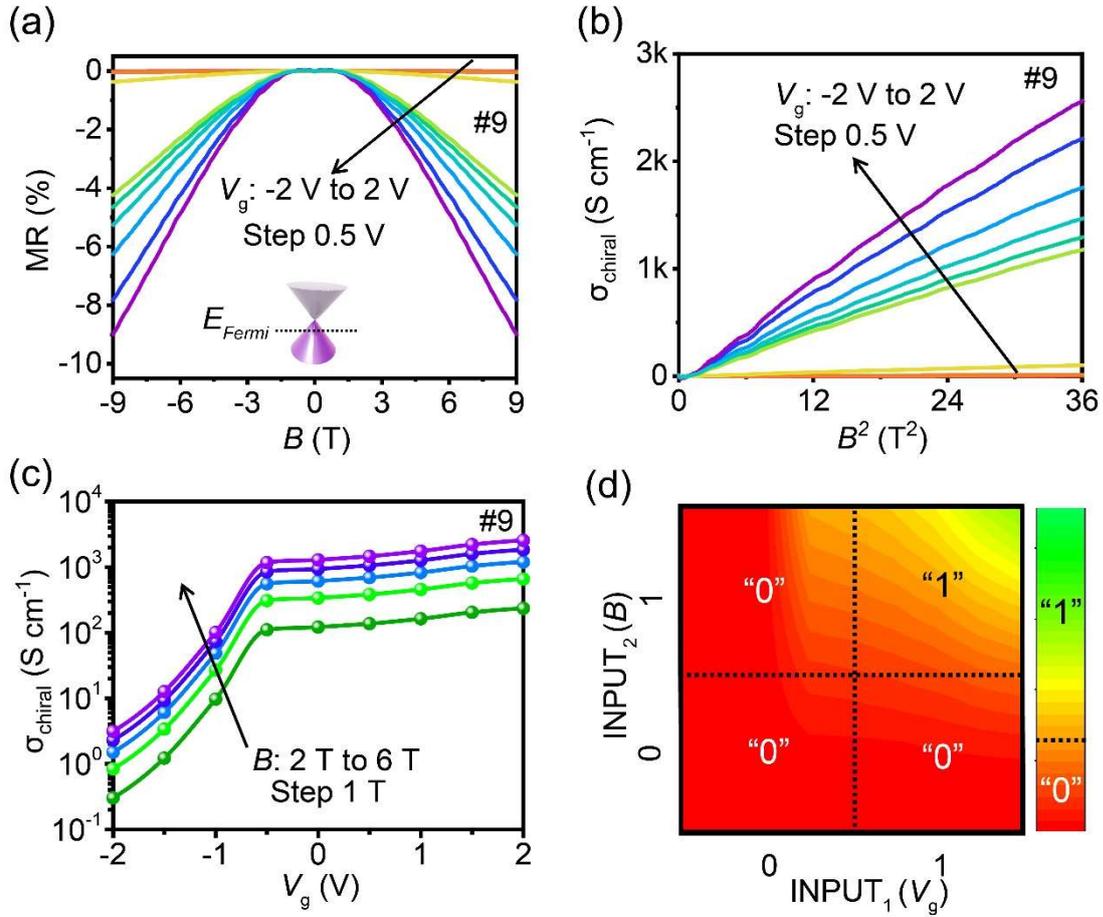

**Figure 3. Chirality-based field-effect PtSe$_2$ device with logic function.** (a) Chiral anomaly based negative MR curves as a function of ***B*** under different $V_g$. The inset shows that $E_{Fermi}$ is below the Dirac point of PtSe$_2$ at $V_g$=0. (b) $\sigma_{chiral}$ as a function of ***B***$^2$ under different $V_g$. (c) $\sigma_{chiral}$ curves as a function of $V_g$ under different ***B***. As the Fermi level moves away from the Dirac point under negative $V_g$, the chiral conductivity decreases remarkably. (d) Demonstration of AND logic gate. INPUT 1 is "0" for ***B*** ≤ 4.5T, and "1" for ***B*** > 4.5T. INPUT 2 is "0" for $V_g$ ≤ 0 V, and "1" for $V_g$ > 0 V. OUTPUT is "0" for $\sigma_{chiral}$ ≤ 780 S cm$^{-1}$, and "1" for $\sigma_{chiral}$ > 780 S cm$^{-1}$.

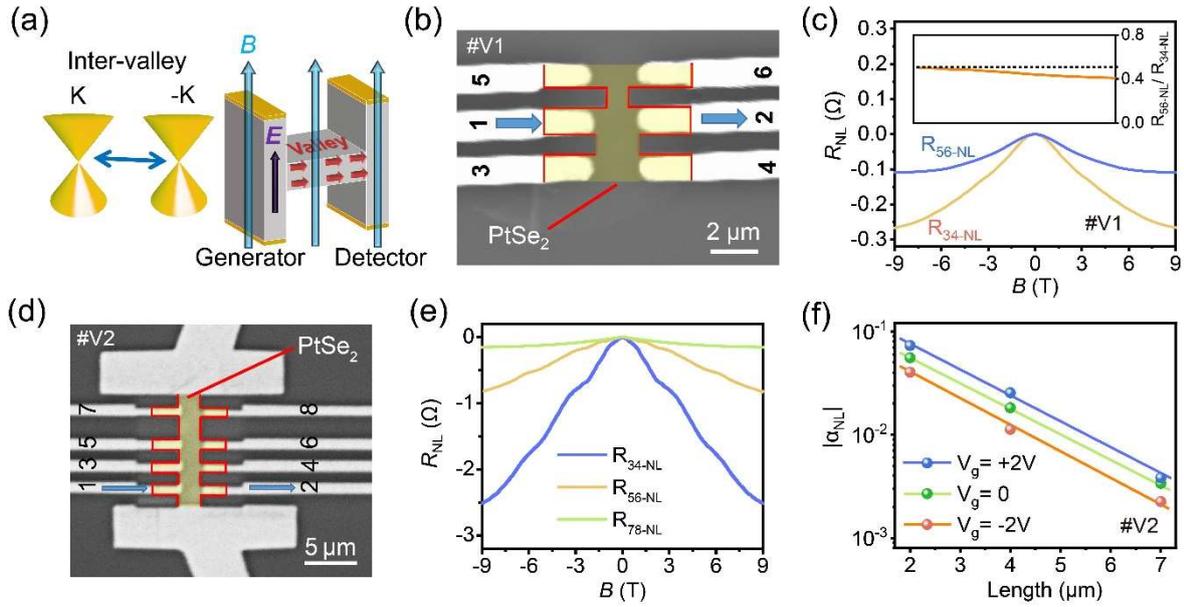

**Figure 4. Chirality-based nonlocal valley transport.** (a) Schematic view of the inter-valley diffusion between different valley positions (±K), compensating the chiral anomaly induced charging pumping process. (b) Scanning electron microscopy of sample #V1. It has the same channel length 2 μm and different channel width of 2 μm (terminal 3-4), and 1 μm (terminal 5-6), respectively. (c) Width-dependent valley transport in sample #V1 under the magnetic field modulation. The inset shows the ratio of $R_{56-NL}/R_{34-NL}$ as a function of $B$ at different temperatures. The dash line corresponds to the theoretical value of 0.50. (d) Scanning electron microscopy of sample #V2. It has the same channel width 2 μm and different channel length from 2 μm (terminal 3-4), 4 μm (terminal 5-6) to 7 μm (terminal 7-8). (e) Length-dependent valley transport sample #V1 under the magnetic field modulation. (f) Electrical modulation of length-dependent valley transport strength through ionic liquid gating. The calculated valley strength $|\alpha_{NL}|$ as a function of diffusion length shows that $|\alpha_{NL}|$ decays exponentially with the increasing of lateral length.